\documentclass[journal]{IEEEtran}
\usepackage{amssymb}
\usepackage{graphicx}
\usepackage{caption}
\usepackage{subfigure}
\usepackage{amsmath}
\usepackage{float}
\usepackage{ragged2e}
\usepackage{color}
\usepackage{epstopdf}
\usepackage{lineno,hyperref}
\usepackage[figuresright]{rotating}
\usepackage{xcolor}
\usepackage{setspace}
\modulolinenumbers[5]

\usepackage[justification=centering]{caption}
\usepackage{booktabs}
\usepackage{multirow}
\usepackage{makecell}
\hypersetup{
colorlinks=true,
linkcolor=black,
citecolor=black,
anchorcolor=black}

\begin{document}
\title{Neuroadaptive Distributed Event-triggered Control of Networked Uncertain Pure-feedback Systems with Polluted Feedback}
\author{Libei~Sun,~Zhirong~Zhang,~Xinjian~Huang,~Xiucai~Huang
\thanks{The authors are with the State Key Laboratory of Power Transmission Equipment System Security and New Technology, Chongqing University, China (e-mail: lbsun@cqu.edu.cn, zhirong.zhang@ntu.edu.sg, huangxinjian@cqu.edu.cn, and hxiucai@cqu.edu.cn).}}
\maketitle

\begin{abstract}
\justifying
This paper investigates the distributed event-triggered control problem for a class of uncertain pure-feedback nonlinear multi-agent systems (MASs) with polluted feedback. Under the setting of event-triggered control, substantial challenges exist in both control design and stability analysis for systems in more general non-affine pure-feedback forms wherein all state variables are not directly and continuously available or even polluted due to sensor failures, and thus far very limited results are available in literature. In this work, a nominal control strategy under regular state feedback is firstly developed by combining neural network (NN) approximating with dynamic filtering technique, and then a NN-based distributed event-triggered control strategy is proposed by resorting to a novel replacement policy, making the non-differentiability issue arising from event-triggering setting completely circumvented. Besides, the sensor ineffectiveness is accommodated automatically without using fault detection and diagnosis unit or controller reconfiguration. It is shown that all the internal signals are semi-globally uniformly ultimately bounded (SGUUB) with the aid of several vital lemmas, while the outputs of all the subsystems reaching a consensus without infinitely fast execution. Finally, the efficiency of the developed  algorithm are verified via numerical simulation.
\justifying
\end{abstract}

\begin{IEEEkeywords}
Distributed event-triggered control, sensor failures, pure-feedback, neural network, multiple agent systems.
\end{IEEEkeywords}
\IEEEpeerreviewmaketitle

\section{Introduction}
\allowdisplaybreaks
With the growing prevalence of networked systems in the era of the Internet of Things (IOT), the event-triggered control has been well recognized as a compelling alternative to traditional time-triggered control, in the aspect of efficiently utilizing shared and limited communication bandwidth and stored energy.
It offers an attractive way for efficient transmissions of the measured signals \cite{astrom2008event}, as the state sensoring, data transmission and control updating occur only if necessary.
Early on, the event-triggered control design has been extensively explored for linear \cite{SEURET201647,heemels2012periodic} and nonlinear systems \cite{tabuada2007event,abdelrahim2015stabilization,abdelrahim2017robust}, among which the results in  \cite{tabuada2007event,abdelrahim2015stabilization,abdelrahim2017robust} are dependent on the input-to-state stability (ISS) property. However, this property is not always guaranteed for nonlinear systems.
Subsequently, this limitation is eliminated successfully by designing event-triggered adaptive backstepping control algorithms in \cite{xing2016event,zhang2021adaptive}.
It is worth underscoring that the works in \cite{SEURET201647}-\cite{zhang2021adaptive} are only
applicable to single nonlinear systems, but not to numerous engineering systems which are actually networked \cite{CHEN2022100004,YANG2022100007}.
For networked multi-agent systems (MASs), the event-triggered control issues under unknown disturbances and actuator failures are investigated in \cite{8474310} and \cite{zhang2018cooperative}, respectively, while the resultant algorithms can only reduce the updating frequency of the actuation signal, that is, the information transmission from sensor to controller remains continuous.
In order to transmit the states under event-triggering setting, considerable solutions have been presented recently. The work in \cite{seyboth2013event} develops a control scheme for multi-agent average consensus with state-triggering execution in the sense that the continuous monitoring of the states of the neighbors involved in \cite{7006773,8319522,8910377} is no longer required.
The idea of developing a distributed adaptive backstepping controller with state-triggering communication is presented in \cite{wang2021adaptive} for norm-form nonlinear systems.
In \cite{sun2022Distributed}, the study on distributed adaptive control under state-triggering action is extended to more general strict-feedback nonlinear systems that are subject to mismatched and nonparametric uncertainties.
Nevertheless, the afore-mentioned methods are built upon that the system states are fully and accurately measurable, which could be not the case in practice since sensor failures are inevitable.

Once sensor failure occurs during the operation of system, the performance of the considered systems under event-triggering setting would be deteriorated since the state information feeding back into the controller is polluted.
It is essential to introduce a suitable compensation mechanism to counteract the effects of sensor failures, and to integrate them with the devised triggering mechanism to ensure that the triggering/sampling error are handled effectively.
Along this direction, some works on nonlinear systems that suffer from sensor failures has emerged recently, see  \cite{wang2019event,cao2019event,zhang2021event} for examples. Specifically, with the aid of auxiliary filters, the authors in \cite{wang2019event} propose an adaptive output-feedback control algorithm against sensor failures under event-triggering setting. Based on the NN approximator, a distributed event-triggered control scheme is developed in \cite{cao2019event} in the presence of sensor failures. In \cite{zhang2021event}, an observer-based decentralized adaptive sensor failure compensation control problem is investigated with event-triggering execution. Whereas, those solutions only consider the control input transmitted over the network while continuous feedback of plant states still exists. Moreover, thus far, in the framework of state-triggered control, studies on non-affine uncertain  pure-feedback nonlinear MASs that are affected by sensor failures remain open, and the relevant issues have not been well solved, although highly desirable.

Enlightened by the above observations, this paper establishes a distributed event-triggered adaptive control strategy for networked uncertain pure-feedback nonlinear MASs with NN approximators, where all state variables are not directly and continuously available or even polluted arising from sensor failures. The main contributions are summarized as follows.
\begin{itemize}
\item [i)]{To overcome the non-differentiability obstacle stemming from the involvement of the intermittent variables, a novel replacement policy is employed in the developed scheme, that is, a nominal control strategy is firstly developed via regular state feedback by using NN approximating and dynamic filtering technique,  and then a NN-based distributed adaptive control scheme with event-triggering setting is constructed by replacing the states in the nominal control strategy with the intermittent ones. Moreover, such replacement ensures the semi-global uniform boundedness of all the internal signals with the aid of several vital lemmas.}

\item [ii)]{The main novelty of handling sensor failures is that the sensor ineffectiveness is accommodated automatically without the utilization of fault detection and diagnosis unit or controller reconfiguration, as opposed to the related state-of-the-art    \cite{zhang2021event,zhai2018output,zhang2018observer}.}

\item [iii)]{To our best knowledge, this is the first solution to the distributed event-triggered control problem for more general uncertain non-affine pure-feedback nonlinear MASs, which intermittently executes state sensoring and actuation signal transmitting in the presence of sensor failures, since almost all of the state-triggered control approaches in the literature are for special kinds of systems  \cite{seyboth2013event,7006773,wang2021adaptive,ZHAN2019104531,wang2020adaptive,long2022output}, like low-order forms, normal forms and  strict-feedback forms.}
\end{itemize}

\section{Preliminaries and Problem Formulation}
\subsection{Gaussian Radial Basis Function (RBF) Networks}
The nonlinear smooth function ${\Psi_{i,k}}\left({{\beta}_{i,k}}\right)\in{\mathcal{R} ^\iota}$ can be approximated by employing the RBFNN \cite{ge2002direct}: 
\begin{flalign}
&\Psi_{i,k}(\beta_{i,k})=\Phi_{i,k}^{T}S_{i,k}(\beta_{i,k}) {\rm{+}}\epsilon_{i,k}{(\beta_{i,k})},\,k=1,\cdots,n& \label{eq:4}
\end{flalign}
where $\beta_{i,k}\in{\mathcal{R}^\iota}
\subset\Omega_{\beta_{i,k}}$ represents the NN input,  $\Omega_{\beta_{i,k}}$ is a compact set, $\Phi_{i,k}\in \mathcal{R}^p$ denotes the weight matrix, which is assumed to belong to a compact set $\Omega_{\Phi_{i,k}}:=\{\left\|\Phi_{i,k}\right\|\le \phi_{i,k0}\}$, with $\phi_{i,k0}$ being some positive constant, $S_{i,k}(\cdot)=\left[S_{i,k1}
(\cdot),\cdots,S_{i,kp}\left(\cdot\right)\right]^{T}\in \mathcal{R}^p$ denotes the basis function vector,    $\epsilon_{i,k}{\left(\beta_{i,k}\right)}\in{\mathcal{R}}$ denotes the approximate error, which satisfy $\left\| {S_{i,k}(\beta_{i,k})} \right\|\le {\bar{s}_{i,km}}$, $\left| {\epsilon_{i,k} \left(\beta_{i,k} \right)} \right| \le {\bar{\epsilon}_{i,km}}$, and ${\bar{s}_{i,km}}$ and ${\bar{\epsilon}_{i,km}}$ are some unknown positive constants. The common choice of  $S_{i,kb}\left(\beta_{i,k}\right)$, $b=1,\cdots,p$, is the following Gaussian function:
\begin{flalign}
&S_{i,kb}\left(\beta_{i,k}\right)= \exp{\left[-\frac{{\left(\beta_{i,k}-\mu_b\right)^{T} \left(\beta_{i,k}-\mu_b\right)}}{{\zeta_b^2}}\right]}& \label{eq:4c}
\end{flalign}
where $\mu_{b}= \left[\mu_{b,1}, \cdots, {\mu_{b,\iota}}\right]^{T}$ and $\zeta_b$ denote the center and the width of the basis function, respectively.

\subsection{Graph Theory}
Consider the graph consisting of $N$ agents represented by $\mathcal{G}=\{\mathcal{V}, \mathcal{E}\}$,  $\mathcal{V}= \{1,\cdots,N\}$ and $\mathcal{E}$ represent a set of vertexes and edges, respectively.
There exists an edge $(i,j)\in\mathcal{E}$ between vertexes $i$ and $j$ if they are able to exchange data.
${N}_i=\{j\in \mathcal{V}|{(i,j)}\in \mathcal{E}\}$ denotes the set of neighbors of vertex $i$ to exchange data. The weighted adjacency matrix is given as ${\mathcal{A}}=[a_{ij}]\in \mathcal{R}^{N \times N}$, if $(j,i)\in \mathcal{E}$,then we can obtain $a_{ij}>0$, and, on the contrary, $a_{ij}=0$. For all $i, j \in \mathcal{V}$, if $a_{ij} = a_{ji}$, it holds that the directed graph $\mathcal{G}$ is undirected. The Laplacian matrix $\mathcal{L}=[l_{ij}]\in \mathcal{R}^{N\times N}$, where $l_{ii}=\sum_{j=1, j \neq i}^{N} a_{i j}$ and $l_{i j}=-a_{i j}, i \neq j$. $\mathcal{D}={\rm{diag}}(d_1,\cdots,d_N)\in \mathcal{R}^{N\times N}$ represents the absolute in-degree matrix, with $d_{i}=\sum_{j\in N_i} a_{ij}$, satisfying $\mathcal{L}=\mathcal{D}-\mathcal{A}$. 

\subsection{System Model}
Consider an uncertain pure-feedback nonlinear MAS composed of $N\,(N\ge{2})$ agents, with the $i$th ($i=1,\cdots,N$) agent modeled as:
\begin{flalign}
{{\dot x}_{i,k}} =\,& {f_{i,k}}\left( {{\check{x}_{i,k}},{x_{i,{k+1}}}} \right),\,k = 1,\cdots ,n - 1 &  \nonumber \\
{{\dot x}_{i,n}} =\,& {f_{i,n}}\left( {{\check{x}_{i,n}},{u_i}}\right)&\nonumber \\
{y_i} =\,& {x_{i,1}} & \label{eq:1}
\end{flalign}
where ${x_{i,k}}\in \mathcal{R}$ is the state, $k=1,\cdots,n$, with ${\check{x}_{i,{k}}}=\left[x_{i,1},\cdots,x_{i,k}\right]^T$, $u_i\in \mathcal{R}$ is the control input, ${f_{i,k}}(\cdot)\in\mathcal{R}$ is the unknown nonaffine continuous function, $k=1,\cdots,n$, and $y_i\in\mathcal{R}$ is the output.

We give the definition of sensor failures as follows.

${\textbf{Definition 1}}$ \cite{WU20051925}.
Let $\chi\left(t\right)\in \mathcal{R}$ and $\chi^f\left(t\right)\in \mathcal{R}$ be the system variable and the output of the sensor, respectively, if $\chi^f\left(t\right)=\eta\left(t\right)\chi\left(t\right)$, $\forall t>\tau_f$, with $0<\eta\left(t\right)\le{1}$, then the sensor is said to have failed at the time $\tau_f$.

In light of \emph{Definition} 1, the potential sensor failures in system (\ref{eq:1}) is modeled as:
\begin{flalign}
&x_{i,k}^f\left(t\right)=\eta_{i}\left(t\right)
x_{i,k}\left(t\right),\,\forall t>\tau_f,\, k=1,\cdots,n& \label{eq:6}
\end{flalign}
where $0<\eta_i\left(t\right)\le{1}$ denotes the fault factor.

${\textbf{Remark 1}}$.
The salient features of the sensor failure model under consideration are twofold.
Firstly, the sensor failure model allows for failures in all plant states, which is more general and challenging than related results in the literature where only the output measurement $y_i$ is destroyed   \cite{zhang2021event,zhai2018output,zhang2018observer}.
Secondly, there are three scenarios that are mainly considered for sensor failures so far, including cases of the partial failures (the effectiveness of the sensor is lost partially), the outage, and the stuck fault. In this work, the partial failures modeled by (\ref{eq:6}) is utilized, that is, although the sensor looses its effectiveness, it is still functional such that $x_{i,k}^{f}$ can always be affected by the plant states $x_{i,k}$.

${\textbf{Remark 2}}$.
The system (\ref{eq:1}) is in a networked uncertain non-affine nonlinear pure-feedback form which is  frequently encountered in various nonlinear control problems for physical systems, such as single-link motor manipulators, hypersonic flight vehicles \cite{7182323}, inverted pendulums, chemical reactors \cite{li2020output} and so on.

\subsection{Control Objectives}
The objective of this work is to propose a distributed event-triggered control scheme for system (\ref{eq:1}) with polluted feedback by using NN approximating, ensuring that all the internal signals are semi-globally uniformly ultimately bounded (SGUUB), with the outputs of all the subsystems reaching a consensus under sensor failures. Meanwhile, infinitely fast execution is precluded.


${\textbf{Assumption 1}}$.
The communication graph $\mathcal{G}$ is undirected and connected.

${\textbf{Assumption 2}}$.
The failure factor $\eta_i\left(t\right)$ in (\ref{eq:6}) satisfies $0<\underline{\eta}_i<\eta_i\left(t\right) \le{1}$, and its first derivative $\dot{\eta}_i\left(t\right)$ meets  $0<\left|\dot{\eta}_i\left(t\right)\right|<\bar{\eta}_{i,d}$, where $\underline{\eta}_i$, $\bar{\eta}_{i,d}>0$ are unknown constants.

${\textbf{Assumption 3}}$ \cite{liu2015adaptive}.
The unknown nonaffine function $f_{i,k}\left(\check{x}_{i,k}, x_{i,k+1}\right)$ satisfies the following inequalities:
\begin{flalign}
\underline{\ell}_{i,k}x_{i,k+1}+\varphi_{i,k1} \leq &\,f_{i,k}\left(\check{x}_{i,k}, x_{i,k+1}\right)-f_{i,k}\left(\check{x}_{i,k}, 0\right)&\nonumber\\
\leq&\, \bar{\ell}_{i,k} x_{i,k+1}+\varphi_{i,k2},\, x_{i,k+1}\ge{0} &\label{eq:v04_2}\\
\underline{\ell}_{i,k}'x_{i,k+1}+\varphi_{i,k1}' \leq &\,f_{i,k}\left(\check{x}_{i,k},x_{i,k+1}\right)
-f_{i,k}\left(\check{x}_{i,k}, 0\right)&\nonumber\\
\leq& \, \bar{\ell}_{i,k}' x_{i,k+1}+\varphi_{i,k2}',\, x_{i,k+1}<0 &\label{eq:v04_22}
\end{flalign}
for $k=1,\cdots,n$, with $x_{i,n+1}=u_i$, where $\underline{\ell}_{i,k}$, $\bar{\ell}_{i,k}$, $\underline{\ell}_{i,k}'$, $\bar{\ell}_{i,k}'$ are some unknown postive constants, $\varphi_{i,k1}$, $\varphi_{i,k2}$, $\varphi_{i,k1}'$ and $\varphi_{i,k2}'$ are some unknown constants.

${\textbf{Remark 3}}$.
To prevent overshadowing the main idea of this work, we consider the case in which the agents are over undirected graphs $\mathcal{G}$, as stated in \emph{Assumption} 1. Nevertheless, it is worth underscoring that the developed algorithm enables the extension to the case of the directed topology by employing the ideas in \cite{7273843,wang2022adaptive}.
In \emph{Assumption} 2, $\eta_i\left(t\right)$ is assumed to satisfy $0<\underline{\eta}_i<\eta_i\left(t\right) \le{1}$, which is quite common in the current literature, see \cite{cao2019event,bounemeur2018indirect} for examples,  and the hypothesis imposed on $\dot{\eta}_i\left(t\right)$ indicates that the growth rate of $\eta_i\left(t\right)$ will not be infinite, which is reasonable in practice.

${\textbf{Remark 4}}$.
In the available methods for dealing with uncertain nonlinear pure-feedback systems \cite{wang2006iss,song2016dealing,wang2011adaptive}, the nonaffine function $f_{i,k}\left(\check{x}_{i,k}, x_{i,k+1}\right)$ is normally assumed to satisfy $0<\underline{\theta}_{i,k} \leq \frac{{\partial f_{i,k}\left(\check{x}_{i,k}, x_{i,k+1}\right)}}{{\partial x_{i,k+1}}}\leq \bar{\theta}_{i,k}$, with $\underline{\theta}_{i,k}$ and $\bar{\theta}_{i,k}$ being some unknown positive constants. However, once ${{\partial f_{i,k}\left(\cdot\right)}}/{{\partial x_{i,k+1}}}$   does not exist (or not strictly positive), those methods are no longer valid. For instance, if there is a dead-zone nonlinearity involved in the model, $f_{i,k}\left(\cdot\right)$ is nondifferentiable with respect to $x_{i,k+1}$.
To eliminate such limitation, in this work a more general assumption is imposed on the nonaffine function $f_{i,k}\left(\cdot\right)$ inspired partly by the ideas in \cite{liu2015adaptive}, as noted in \emph{Assumption} 3. Clearly, \emph{Assumption} 3 satisfies the condition in existing results, while remaining valid even if ${{\partial f_{i,k}\left(\cdot\right)}}/{{\partial x_{i,k+1}}}$  does not exist (or not strictly positive). Consequently, the proposed solution is more general than the existing ones.

For establishing the stability results, we introduce the following lemma.

${\textbf{Lemma 1}}$ \cite{horn2012matrix}.
Let the graph $\mathcal{G}$ be undirected and connected, it holds that $\mathcal{L}$ has a simple eigenvalue of 0, with all other eigenvalues being positive, that is, $\lambda_1 = 0$ and $\lambda_i>0$, $i=2,\cdots,N$.

\subsection{System Transformation}
For notation conciseness, we denote $h_{i,k}\left(\check{x}_{i,k}, x_{i,k+1}\right)=f_{i,k}\left(\check{x}_{i,k}, x_{i,k+1}\right)-f_{i,k}\left(\check{x}_{i,k}, 0\right)$, $k=1,\cdots,n$. It can be derived from (\ref{eq:v04_2}) and (\ref{eq:v04_22}) that there exist functions $\varpi_1\left(\check{x}_{i,k+1}\right)$ and $\varpi_2\left(\check{x}_{i,k+1}\right)$ that take values in [0,1], satisfying
\begin{flalign}
h_{i,k}&\left(\check{x}_{i,k}, x_{i,k+1}\right)=\left(1-\varpi_1
\left(\check{x}_{i,k+1}\right)\right)
\left(\underline{\ell}_{i,k}
x_{i,k+1}+\varphi_{i,k1}\right)&\nonumber\\
&+\varpi_1\left(\check{x}_{i,k+1}\right)\left(\bar{\ell}_{i,k}
x_{i,k+1}+\varphi_{i,k2}\right),x_{i,k+1}\ge{0}&\label{eq:v04_1}\\
h_{i,k}&\left(\check{x}_{i,k}, x_{i,k+1}\right)=\left(1-\varpi_2
\left(\check{x}_{i,k+1}\right)\right)\left(\underline{\ell}_{i,k}'
x_{i,k+1}+\varphi_{i,k1}'\right)&\nonumber\\
&+\varpi_2\left(\check{x}_{i,k+1}\right)\left(\bar{\ell}_{i,k}'
x_{i,k+1}+\varphi_{i,k2}'\right),x_{i,k+1}<{0}.&\label{eq:v04_1a}
\end{flalign}
To facilitate the analysis, the following functions are introduced
\begin{flalign}
g_{i,k}(\check{x}_{i,k+1})= \begin{cases}
\left(1-\varpi_1\left(\check{x}_{i,k+1}
\right)\right)\underline{\ell}_{i,k}\\
\qquad +\varpi_1
\left(\check{x}_{i,k+1}\right)\bar{\ell}_{i,k},x_{i,k+1} \geq 0 \\
\left(1-\varpi_{2}\left(\check{x}_{i,k+1}\right)\right) \underline{\ell}_{i,k}^{\prime}\\
\qquad +\varpi_{2}
\left(\check{x}_{i,k+1}\right)\bar{\ell}_{i,k}^{\prime},
x_{i,k+1}<0\end{cases} \label{eq:v04_1b} 
\end{flalign}
\begin{flalign}
d_{i,k}\left(\check{x}_{i,k+1}\right)= \begin{cases}\left(1-\varpi_{1}\left(\check{x}_{i,k+1}
\right)\right)\varphi_{i,k1}\\
\quad +\varpi_{1}\left(\bar{x}_{i,k+1}
\right) \varphi_{i,k2},x_{i,k+1} \geq 0 \\
\left(1-\varpi_{2}\left(\check{x}_{i,k+1}\right)\right) \varphi_{i,k1}'\\
\quad +\varpi_{2}\left(\check{x}_{i,k+1}\right) \varphi_{i,k2}', x_{i,k+1}<0.
\end{cases}\label{eq:v04_1c}
\end{flalign}
for $k=1,\cdots,n$. This, together with (\ref{eq:v04_1}), (\ref{eq:v04_1a}) and the definition of $h_{i,k}\left(\cdot\right)$, results in
\begin{flalign}
f_{i,k}\left(\check{x}_{i,k},x_{i,k+1}\right)
=&\,\,g_{i,k}\left(\check{x}_{i,k+1}\right)x_{i,k+1}
+f_{i,k}\left(\check{x}_{i,k},0\right)&\nonumber\\
&+d_{i,k}(\check{x}_{i,k+1})
&\label{eq:vv02}
\end{flalign}
with $0<\,\underline{g}_{i,k}\leq g_{i,k}\left(\check{x}_{i,k+1}\right)\leq \bar{g}_{i,k}$, and $0\leq\,\left|d_{i,k}\left
(\check{x}_{i,k+1}\right)\right| \leq \bar{d}_{i,k}$, $k=1,\cdots,n$, where
$\underline{g}_{i,k}=\min\{\underline{\ell}_{i,k}, \bar{\ell}_{i,k},\underline{\ell}_{i,k}', \bar{\ell}_{i,k}'\}$, $\bar{g}_{i,k}=\max\{\underline{\ell}_{i,k},
\bar{\ell}_{i,k},\underline{\ell}_{i,k}', \bar{\ell}_{i,k}'\}$ and $\bar{d}_{i,k}=\max\{\left|\varphi_{i,k1}\right|
+\left|\varphi_{i,k2}\right|, |\varphi_{i,k1}'|
+|\varphi_{i,k2}'|\}$. In light of the definition of $g_{i,k}(\cdot)$ in (\ref{eq:v04_1b}) and recalling (\ref{eq:1}), it is clear that $\dot{g}_{i,k}(\cdot)$ only depends on states $x_{i,k}$. As $f_{i,k}(\cdot)$, ${\varpi_{1}(\cdot)}$ and ${\varpi_{2}(\cdot)}$ are smooth functions, $\dot{g}_{i,k}(\cdot)$, $k=1, \cdots, n$, is therefore bounded, that is, $\left|\dot{g}_{i,k}(\cdot)\right| \leqslant \bar{g}_{i,d}$, where $\bar{g}_{i,d}>0$ are some unknown constant.

With the above analysis and (\ref{eq:vv02}), we can rewrite system (\ref{eq:1}) as
\begin{flalign}
{{\dot x}_{i,k}} =\,& {g_{i,k}}(\check{x}_{i,k+1})
{x_{i,{k+1}}}{\rm{+}}{f_{i,k}}({{\check{x}_{i,{k}}},0})
+d_{i,k}\left(\check{x}_{i,k+1}\right) &  \nonumber \\
{{\dot x}_{i,n}} =\,& {g_{i,n}}\left(\check{x}_{i,n+1} \right){u_{i}}+{f_{i,n}}\left( {{\check{x}_{i,{n}}},0}
\right)+d_{i,n}\left(\check{x}_{i,n+1}\right)
&\nonumber \\
{y_i} =\,& {x_{i,1}}. & \label{eq:1a}
\end{flalign}
for $k = 1,\cdots ,n -1$.


\section{Nominal Control Design}
To promote the design of the distributed adaptive control with event-triggered setting, a nominal control scheme is firstly developed in this section by combining NN approximating with dynamic filtering technique.

To proceed, we define the following coordination transformation $z_{i,k}(k=1,\cdots,n)$ and consensus error $e_i$:
\begin{flalign}
{z_{i,1}}=\,& x_{i,1}^f & \label{eq:7}\\
{z_{i,k}}=\,& {x_{i,k}^f} - {\alpha _{i,kf}}& \label{eq:8}\\
{e_i}=\,& \sum\limits_{j = 1}^N {{a_{ij}}} \left( {{x_{i,1}^f} - {x_{j,1}^f}} \right)& \label{eq:9}
\end{flalign}
for $k = 2, \cdots ,n$, where $\alpha_{i, kf}$ denotes the output of the following filter:
\begin{flalign}
&{\xi _{i,k}}{\dot \alpha _{i,kf}} +{\alpha _{i,kf}} ={\alpha _{i,k - 1}},\,{\alpha _{i,kf}}\left( 0 \right) = {\alpha _{i,k - 1}}\left( 0 \right)& \label{eq:10}
\end{flalign}
where ${\xi_{i,k}}$ is some positive design parameter, $\alpha_{i,k-1}$ denotes a virtual control that serves as the input of (\ref{eq:10}). For convenience of stability analysis, we further define, for $k=2,\cdots,n$
\begin{flalign}
&{\Theta_{i,k}} = {\alpha _{i,kf}} - {\alpha _{i,k - 1}}& \label{eq:11}
\end{flalign}

We design the following nominal control scheme through regular state feedback:
\begin{flalign}
{\alpha _{i,1}} =&\, - {\delta_1}{e_i} -\left
(\gamma_{i,1}+1\right){z_{i,1}}- \hat \Phi_{i,1}^T
{S _{i,1}}\left( {{\beta_{i,1}}}\right)&\label{eq:12}\\
{\alpha_{i,k}} =&\,-\left(\gamma_{i,k}+1\right)
{z_{i,k}} - {z_{i,k - 1}} - \hat \Phi_{i,k}^T{S_{i,k}}\left( {{\beta _{i,k}}} \right)&\label{eq:13}\\
{u_i}=&\,\,{\alpha_{i,n}}& \label{eq:14}
\end{flalign}
for $k=2,\cdots,n$, where $\delta_1>0$ and $\gamma_{i,k}>0$ are the design parameters, $k=1,\cdots,n$. The updating law of ${{\hat \Phi}_{i,k}}$ is as follows:
\begin{flalign}
&{{\dot {\hat \Phi}}_{i,k}}= Proj\left[{\Lambda _{i,k}}{S_{i,k}}\left({\beta }_{i,k} \right){z}_{i,k}\right], \, k=1,\cdots,n& \label{eq:15}
\end{flalign}
with ${\hat \Phi}_{i,k}\left(0\right)
\in\Omega_{\Phi_{i,k}}$, where $\hat{\Phi}_{i,k}$ denotes the estimate of ${\Phi}_{i,k}$, satisfying ${\tilde \Phi}_{i,k}={\Phi}_{i,k}- \hat{\Phi}_{i,k}$, $\Lambda_{i,k}$ denotes some matrix that is positive definite, $Proj\left(\cdot\right)$ denotes a projection operator to ensure that ${\hat \Phi}_{i,k}$ belongs to compact set $\Omega_{\hat{\Phi}_{i,k}}:=\left\{\| {{{\hat \Phi}_{i,k}}} \| \le {\phi_{i,k0}} + \iota \buildrel \Delta \over = {{\bar \phi}_{i,k}}\right\}$, which further indicates that $\| {{{\tilde \Phi}_{i,k}}}\|\le 2{\phi_{i,k0}} + \iota  \buildrel \Delta \over = {\phi_{i,km}}$, where $\iota>0$ is an arbitrary constant, and ${{\bar \phi}_{i,k}}, \phi_{i,km}>0$ are some unknown constants.

We present the main results for nominal control scheme of system (\ref{eq:1}) with polluted feedback.

${\textbf{Theorem 1}}$.
Consider an uncertain non-affine pure-feedback nonlinear MAS of $N$ agents (\ref{eq:1}) with   undirected topology. If \emph{Assumptions} 1-3 hold, applying distributed controller (\ref{eq:14}) and the adaptive law (\ref{eq:15}), then
\begin{itemize}
\item [i)]{All the internal signals are SGUUB.}
\item [ii)]{The outputs of all the subsystems reach a consensus in the presence of sensor failures, and the upper bound of ${\left\|y\right\|_{[0, T]}}$ can be decreased by selecting the design parameters appropriately.}
\end{itemize}

\textbf{Proof}. See \emph{Appendix} A.

${\textbf{Remark 5}}$.
The major design difficulties in dealing with sensor failures are twofold: i) all state variables are not directly available for feedback design owing to sensor failures; and ii) the sensor failures under consideration (as seen in (\ref{eq:6})) are more general and challenging than existing results where only the output measurement $y_i$ is unavailable for controller design   \cite{zhang2021event,zhai2018output,zhang2018observer}. To address such issues, the neuroadaptive fault-tolerant method is applied in this work that allows the sensor ineffectiveness to be accommodated automatically without using fault detection and diagnosis unit or controller reconfiguration.

\section{Event-triggered Control Design}
This section proposes a distributed adaptive control scheme with event-triggering setting based on the nominal control scheme, which is capable of coping with the non-differentiability issue and sensor failures simultaneously by resorting to a novel replacement policy and NN approximating.

{\subsection{Design of Triggering Conditions}}
Let $t_{k,p}^i$, $t_{k,p}^j$ and $t_{u,p}^i$ be the sequences of transmissions of $x_{i,k}^f$, $x_{j,k}^f$ and $v_i$, respectively, $i,j=1, \ldots, N$, $k=1, \ldots, n$, $p = 0,1,2, \cdots$.
At each transmission instant $t_{k,p}^i$, the local state measurement $x_{i,k}^f$ is sent to the controller, $t_{k,p}^j$ is the transmission instant at which agent $j$ broadcasts its $k$th state information to agent $i$, and the input measurement $v_i$ is broadcasted to the actuators at $t_{u,p}^i$. This indicates that the instantaneous information of local state $x_{i,k}^f$, neighboring subsystem state $x_{j,k}^f$ and control input $v_i$ are updated only at the time instants $t_{k,p}^i$, $t_{k,p}^j$ and $t_{u,p}^i$, respectively, then the following equalities hold:
\begin{flalign}
{{\bar x}_{i,k}^f}\left( t \right) =\,&{x_{i,k}^f}\left( {t_{k,p}^i} \right),\,\forall t \in [t_{k,p}^i,t_{k,p + 1}^i)& \label{eq:37}\\
{{\bar x}_{j,k}^f}\left( t \right) =\,&{x_{j,k}^f}\left( {t_{k,p}^j} \right),\,\forall t \in [t_{k,p}^j,t_{k,p + 1}^j)& \label{eq:38}\\
{u_i}\left( t \right) =\,&{v_i}\left( {t_{u,p}^{i}} \right),\,\,\forall t \in [t_{u,p}^{i},t_{u,p + 1}^{i}).& \label{eq:39}
\end{flalign}

Going forward, we design the triggering conditions as follows:
\begin{flalign}
t_{k,p + 1}^i=& \inf \left\{ {t > t_{k,p}^i,\left| {{x_{i,k}^f}\left( t \right) - {{\bar x}_{i,k}^f}\left( t \right)} \right| > \Delta x_{i,k}} \right\} & \label{eq:40}\\
t_{k,p + 1}^j=& \inf \left\{ {t > t_{k,p}^j,\left| {{x_{j,k}^f}\left( t \right) - {{\bar x}_{j,k}^f}\left( t \right)} \right| > \Delta x_{j,k}} \right\} & \label{eq:41}\\
t_{u,p + 1}^i=& \inf \left\{ {t > t_{u,p}^i,\left| {{v_{i}}\left( t \right) - {{u}_{i}}\left( t \right)} \right| > \Delta u_{i}} \right\} & \label{eq:42}
\end{flalign}
where $t_{k,0}^i$, $t_{k,0}^j$ and $t_{u,0}^{i}$ represent the first instant when (\ref{eq:40})-(\ref{eq:42}) are fulfilled, respectively, and $\Delta x_{i,k}>0$, $\Delta x_{j,k}>0$ and $\Delta u_{i}>0$ are the designed triggering thresholds. 

{\subsection{Controller Design}}
Under intermittent state feedback circumstances, the coordination transformation $z_{i,k}\,(k=1,\cdots,n)$ and the consensus error $e_i$ in (\ref{eq:7})-(\ref{eq:9}) are converted as:
\begin{flalign}
{\bar{z}_{i,1}}=\,& {\bar{x}}_{i,1}^f & \label{eq:43}\\
{\bar{z}_{i,k}}=\,& {\bar{x}_{i,k}^f} - {\bar{\alpha} _{i,kf}} & \label{eq:44}\\
{\bar{e}_i}=\,& \sum\limits_{j = 1}^N {{a_{ij}}} \left( {{\bar{x}_{i,1}^f} - {\bar{x}_{j,1}^f}} \right).& \label{eq:45}
\end{flalign}
for $k = 2, \cdots ,n - 1$.
Next, we present the following distributed event-triggered control strategy via intermittent state feedback:
\begin{flalign}
{{\bar {\alpha}}_{i,1}} =&\, -{\delta _1}{{\bar e}_i}- \left({\gamma_{i,1}}+1\right){{\bar z}_{i,1}} - \hat \Phi_{i,1}^T{S_{i,1}}\left( {{{\bar \beta }_{i,1}}} \right) & \label{eq:46}\\
{{\bar \alpha }_{i,k}}=&\, -\left({\gamma_{i,k}}+1\right) {{\bar z}_{i,k}} - {{\bar z}_{i,k - 1}}- \hat \Phi_{i,k}^T{S_{i,k}}\left( {{{\bar \beta }_{i,k}}} \right)& \label{eq:47}\\
{v_i} =&\,\, {{\bar \alpha }_{i,n}}& \label{eq:48}
\end{flalign}
for $k=2,\cdots,n$, where $\delta_1>0$ and $\gamma_{i,k}>0$ are the design parameters, $k=1,\cdots,n$. The updating law of ${{\hat \Phi}_{i,k}}$ is as follows:
\begin{flalign}
&{{\dot {\hat \Phi}}_{i,k}} = Proj\left[{\Lambda _{i,k}}{S_{i,k}}\left({\bar \beta }_{i,k} \right){\bar z}_{i,k}\right],\,k=1,\cdots,n& \label{eq:49}
\end{flalign}
with ${\hat \Phi}_{i,k}\left(0\right)
\in\Omega_{\phi_{i,k}}$, where $Proj\left(\cdot\right)$ is a projection operator defined previously, and $\Lambda_{i,k}$ denotes some matrix that is positive definite.

To proceed, we establish several vital lemmas.

${\textbf{Lemma 2}}$ \cite{sun2022Distributed}.
Let ${S_{i,k}}\left( {{\beta _{i,k}}} \right)$ be the basis function vector defined in (\ref{eq:4}), $k=1,\cdots,n$, it holds that
\begin{flalign}
&\left\| {{S_{i,k}}\left( {{\beta _{i,k}}} \right) - {S_{i,k}}\left( {{{\bar \beta }_{i,k}}} \right)} \right\| \le \Delta {s_{i,k}}&\label{eq:52_a}
\end{flalign}
where $\Delta {s_{i,k}}>0$ is a constant relies on the design parameters $\Delta x_{i,k}$ and $\zeta_b$, $b=1,\cdots,p$.

${\textbf{Lemma 3}}$. Let $z_{i,k}$ and $\alpha_{i,k}$ be defined in (\ref{eq:7}), (\ref{eq:8}), (\ref{eq:12}), (\ref{eq:13}), respectively, then the following inequalities hold
\begin{flalign}
\left|{{z_{i,k}}-{{\bar z}_{i,k}}}\right|\le &\, \Delta {z_{i,k}} & \label{eq:51} \\
\left| {{\alpha _{i,k}} {\rm{-}} {{\bar \alpha }_{i,k}}} \right| \le &\, \Delta {\alpha _{i,k}} &\label{eq:52}
\end{flalign}
for $k=1,\cdots,n$, where $\Delta {z_{i,k}}>0$ and $\Delta {\alpha_{i,k}}>0$ are constants that rely on the design parameters $\Delta x_{i,k}$, $\Delta x_{j,k}$, $\gamma_{i,k}$, $d_i$ and $\zeta_b$, $b=1,\cdots,p$.

{\textbf{Proof}}.
From (\ref{eq:9}) and (\ref{eq:45}), it follow that
\begin{flalign}
{\left|{{e_i}-{\bar e}_i}\right|}
\le \,& {d_i}\left(\Delta {x_{i,1}} + \Delta x_{j,1} \right)\, \buildrel \Delta \over =\Delta {e_i}.& \label{eq:a1}
\end{flalign}
As observed from (\ref{eq:7}) and (\ref{eq:43}), we can obtain
\begin{flalign}
&\left| {{z_{i,1}} - {{\bar z}_{i,1}}} \right|
\le\Delta {x_{i,1}}\, \buildrel \Delta \over = \Delta {z_{i,1}}. & \label{eq:a2}
\end{flalign}
In accordance with (\ref{eq:12}) and (\ref{eq:46}), it can be derived as
\begin{flalign}
\left| {{\alpha _{i,1}} - {{\bar \alpha }_{i,1}}} \right| \le \,&{\delta _1} \left|\bar{e}_i-e_i\right| + \left({\gamma_{i,1}}+1\right) \left|{\bar{z}_{i,1}}-z_{i,1}
\right|&\nonumber\\
&+ \left| {\hat \Phi_{i,1}^T\left( {{S_{i,1}}\left( {{{\bar \beta }_{i,1}}} \right) - {S_{i,1}}\left( {{\beta _{i,1}}} \right)} \right)} \right|& \nonumber\\
\le\, & {\delta _1}\Delta {e_i} + \gamma_{i,1} \Delta {z_{i,1}}+ \Delta {z_{i,1}}&\nonumber\\
&+\bar{\phi}_{i,1}\Delta{s_{i,1}}\buildrel \Delta \over = \Delta {\alpha_{i,1}} &\label{eq:a3}
\end{flalign}
For simplicity of presentation, we denote $\tilde{\alpha}_{i,kf}={\alpha}_{i,kf}
-{\bar{\alpha}}_{i,kf}$, $k=2,\cdots,n$.
By recalling (\ref{eq:10}), one can obtain
\begin{flalign}
\dot{\tilde{\alpha}}_{i,kf}
=& -\frac{1}{\xi_{i,k}}\tilde{\alpha}_{i,kf}
+\frac{1}{\xi_{i,k}} \left({\alpha _{i,k - 1}}-{\bar{\alpha}_{i,k - 1}}\right).
&\label{eq:a4}
\end{flalign}
By integrating both sides of (\ref{eq:a4}), it is not difficult to derived that
\begin{flalign}
\left|{\tilde{\alpha}}_{i,kf}\right| \le & \left|\tilde{\alpha}_{i,kf}\left(0\right)\right|
+e^{-\frac{1}{\xi_{i,k}}t}\Delta \alpha_{i,{k-1}}\left(e^{\frac{1}{\xi_{i,k}}t}-1 \right)&\nonumber\\
\le&\left|\tilde{\alpha}_{i,kf}\left(0\right)\right|+\Delta \alpha_{i,{k-1}} \buildrel \Delta \over =\Delta \alpha_{i,kf} &\label{eq:a5_a}
\end{flalign}
for $k=2,{\rm{\cdots}},n$. Using (\ref{eq:8}) and (\ref{eq:44}), we have
\begin{flalign}
&\left| {{z_{i,2}} - {{\bar z}_{i,2}}} \right|
\le\Delta {x_{i,2}} + \Delta {\alpha _{i,2f}}\buildrel \Delta \over = \Delta {z_{i,2}}.& \label{eq:a6}
\end{flalign}
This, together with (\ref{eq:13}) and (\ref{eq:47}), results in
\begin{flalign}
\left| {{\alpha _{i,2}}-{{\bar \alpha }_{i,2}}} \right| \le \,& {\gamma_{i,2}}\Delta {z_{i,2}}+\Delta {z_{i,2}}+ \Delta {z_{i,1}}&\nonumber\\
&+\bar{\phi}_{i,2}\Delta{s_{i,2}}
\, \buildrel \Delta \over =\Delta {\alpha_{i,2}} &\label{eq:a7}
\end{flalign}
In the same vein, it holds that
\begin{flalign}
&\left| {{z_{i,k}} - {{\bar z}_{i,k}}} \right| \le\Delta {x_{i,k}} {\rm{+}} \Delta {\alpha _{i,kf}}
\, \buildrel \Delta \over = \Delta {z_{i,k}},k=3,\cdots,n& \label{eq:a8}
\end{flalign}
\begin{flalign}
\left| {{\alpha _{i,k}} - {{\bar \alpha }_{i,k}}} \right| \le\,& {\gamma_{i,k}}\Delta {z_{i,k}}+ \Delta {z_{i,k}}+ \Delta {z_{i,{k-1}}}&\nonumber\\
&+ \bar{\phi}_{i,k}\Delta{s_{i,k}}\buildrel \Delta \over = \Delta {\alpha_{i,k}},k=3,\cdots,n&\label{eq:a9}
\end{flalign}
Thus we can derive from (\ref{eq:a2}), (\ref{eq:a3}), (\ref{eq:a6})-(\ref{eq:a9}) that \emph{Lemma} 3 holds.  $\hfill{\blacksquare}$

With those lemmas, we state the main results for distributed event-triggered control of system (\ref{eq:1}) with polluted feedback.

${\textbf{Theorem 2}}$.
Consider an uncertain non-affine pure-feedback nonlinear MAS of $N$ agents (\ref{eq:1}) with undirected topology, if \emph{Assumptions} 1-3 hold, applying distributed controller (\ref{eq:48}), adaptive law (\ref{eq:49}) and the event-triggering mechanism (\ref{eq:40})-(\ref{eq:42}), then
\begin{itemize}
\item [i)]{All the internal signals are SGUUB.}
\item [ii)]{The outputs of all the subsystems reach a consensus in the presence of sensor failures, and the upper bound of ${\left\|y\right\|_{[0, T]}}$ can be decreased by selecting the design parameters appropriately.}
\item [iii)]{The Zeno solutions are ruled out.}
\end{itemize}

\textbf{Proof}. The following recursive control design are performed to derive the results in \emph{Theorem} 2.

$\textbf{Step 1}$:
Define a Lyapunov function ${V_1}= \sum\nolimits_{i = 1}^N\frac{1}{2g_{i,1}}z_{i,1}^2 + \sum\nolimits_{i = 1}^N \frac{1}{2}\tilde \Phi_{i,1}^T\Lambda_{i,1}^{ - 1}{{\tilde \Phi}_{i,1}} + \sum\nolimits_{i = 1}^N \frac{1}{2}\Theta_{i,2}^2$. In view of (\ref{eq:6}), (\ref{eq:1a}), (\ref{eq:7}), (\ref{eq:8}) and (\ref{eq:11}), $\dot{V}_1$ can be expressed as
\begin{flalign}
\dot{V}_{1} =&-\sum\limits_{i = 1}^N \frac{\dot{g}_{i,1}}{2g_{i,1}^2}z_{i,1}^2
-\sum\limits_{i = 1}^N\tilde \Phi_{i,1}^T\Lambda _{i,1}^{ - 1}{{\dot{\hat \Phi}}_{i,1}} + \sum\limits_{i = 1}^N\Theta_{i,2}\dot{\Theta}_{i,2} &\nonumber\\
&+\sum\limits_{i = 1}^N{z}_{i,1}\left(z_{i,2}+\Theta_{i,2}
+\alpha_{i,1}+\Psi_{i,1}\left( {{\beta _{i,1}}} \right)\right)&\label{eq:17a}
\end{flalign}
where $\Psi_{i,1}\left( {{\beta _{i,1}}} \right)=\frac{1}{g_{i,1}}\left({\dot{\eta}}_i
{x_{i,1}}+{\eta}_i f_{i,1}\left(x_{i,1},0\right)+{\eta}_i d_{i,1}\right)$ can be approximated by utilizing the RBFNN, namely,  $\Psi_{i,1}(\beta_{i,1})=\Phi_{i,1}^{T}S_{i,1}(\beta_{i,1}) +\epsilon_{i,1}(\beta_{i,1})$, with $\beta _{i,1}=[x_{i,1}, \eta_i,\dot{\eta}_i]^{T} \in{\Omega_{\beta_{i,1}}}$.
Applying \emph{Lemma} 1, it is obtained that
\begin{flalign}
-\sum\limits_{i = 1}^N {\delta _1}{e_i}{z_{i,1}}
\le& -{\delta _1}\sum\limits_{i = 1}^N\sum\limits_{j = 1}^N {{a_{ij}}} \left( {{x_{i,1}^f} - {x_{j,1}^f}} \right)x_{i,1}^f&\nonumber\\
\le & -\delta_1M^{T}\mathcal{L}M
\le -\delta_1\underline {\lambda}
\left\|y\right\|^2&\label{eq:a17_a}
\end{flalign}
where $M=\eta y$, with $\eta={\rm{diag}}(\eta_i)$, $y=\left[y_1,\cdots,y_N\right]^T$, and $\underline {\lambda}$ is the smallest eigenvalue of $M^{T}\mathcal{L}M$. Synthesizing (\ref{eq:12}), (\ref{eq:17a}), (\ref{eq:a17_a}) and adopting Young's inequality, it can be determined that
\begin{flalign}
\dot{V}_{1} \le& -\delta_1\underline{\lambda}\left\|y\right\|^2
-\sum\limits_{i = 1}^N \jmath_{i,1}z_{i,1}^2- \sum\limits_{i =1}^N\xi_{i,2}^{*}\Theta_{i,2}^2&\nonumber\\
&+ \sum\limits_{i = 1}^N {\tilde \Phi_{i,1}^T {{S_{i,1}}\left( {{\beta _{i,1}}} \right){z_{i,1}} - \sum\limits_{i = 1}^N\tilde \Phi_{i,1}^T\Lambda_{i,1}^{-1}{{\dot {\hat \Phi}}_{i,1}}}}&\nonumber\\
&+\sum\limits_{i = 1}^N{z}_{i,1}z_{i,2}+\sum\limits_{i = 1}^N\varsigma_{i,1}^2+\Sigma_1&\label{eq:53}
\end{flalign}
where $\jmath_{i,1}=\gamma_{i,1}-\frac{\bar{g}_{i,d}}
{2\underline{g}_{i,1}^2}>0$ by choosing $\gamma_{i,1}$ large enough, $\xi_{i,2}^*>0$ is the design parameter, with $\frac{1}{\xi_{i,2}}\ge \xi_{i,2}^*+ \frac{3}{4}$, ${\varsigma_{i,1}}=-{\dot \alpha _{i,1}}$, ${\Sigma_1} =\sum\nolimits_{i =1}^N\frac{1}{2}\bar{\epsilon}_{i,1m}^2$.
and the fact that $-\left(\gamma_{i,1}+
\frac{\dot{g}_{i,1}}{2{g}_{i,1}^2}\right)\le -\left(\gamma_{i,1}-\frac{\bar{g}_{i,d}}
{2\underline{g}_{i,1}^2}\right)$ is utilized.


$\textbf{Step \emph{k}}$ $(k=2,\cdots,n-1)$:
Define the Lyapunov function ${V_k} = {V_{k - 1}} + \sum\nolimits_{i = 1}^N \frac{1}{2g_{i,k}}z_{i,k}^2 + \sum\nolimits_{i = 1}^N \frac{1}{2}\tilde \Phi_{i,k}^T\Lambda_{i,k}^{-1}{{\tilde \Phi}_{i,k}}+ \sum\nolimits_{i = 1}^N \frac{1}{2}\Theta_{i,k + 1}^2$. Applying (\ref{eq:6}), (\ref{eq:1a}), (\ref{eq:8}) and (\ref{eq:11}), the derivative of ${V}_{k}$ is derived as
\begin{flalign}
\dot{V}_{k} \le& -\delta_1\underline{\lambda}
\left\|y\right\|^2-\sum\limits_{i = 1}^N\sum\limits_{q = 1}^{k-1} \jmath_{i,q}z_{i,q}^2- \sum\limits_{i = 1}^N\sum\limits_{q =1}^{k-1}\xi_{i,{q+1}}^{*}
\Theta_{i,{q+1}}^2&\nonumber\\
&-\sum\limits_{i = 1}^N \frac{\dot{g}_{i,k}}{2g_{i,k}^2}z_{i,k}^2-\sum\limits_{i = 1}^N\tilde \Phi_{i,k}^T\Lambda_{i,k}^{ - 1}{{\dot{\hat \Phi}}_{i,k}}+\sum\limits_{i = 1}^N{z}_{i,k}{z}_{i,k+1} &\nonumber\\
&+\sum\limits_{i =1}^N{z}_{i,k}
({\Theta}_{i,{k+1}}+\alpha_{i,k}+z_{i,k-1}+\Psi_{i,k}( {{\beta _{i,k}}}))&\nonumber\\
&+ \sum\limits_{i = 1}^N \sum\limits_{q = 1}^{k-1} {\tilde \Phi_{i,q}^T {{S_{i,q}}\left( {{\beta _{i,q}}} \right){z_{i,q}} -\sum\limits_{i = 1}^N\sum\limits_{q = 1}^{k-1} \tilde \Phi_{i,q}^T\Lambda _{i,q}^{-1}{{\dot {\hat \Phi}}_{i,q}}} }&\nonumber\\
&+\sum\limits_{i = 1}^N\Theta_{i+1}\dot{\Theta}_{i+1}
+\sum\limits_{i =1}^N\sum\limits_{q=1}
^{k-1}\varsigma_{i,q}^2+\Sigma_{k-1}
&\label{eq:27a}
\end{flalign}
where ${\Psi_{i,k}}\left( {{\beta _{i,k}}} \right)= \frac{1}{g_{i,k}}({\dot{\eta}}_i{x_{i,k}}+{\eta}_i f_{i,k}\left(\check{x}_{i,k},0\right)+{\eta}_i d_{i,k}-{\dot{\alpha} _{i,kf}})$ is approximated making use of the RBFNN, with $\beta _{i,k}=[\check{x}_{i,k}
,{x}_{j,1},\eta_i,\dot{\eta}_i]^{T} \in{\Omega_{{\beta _{i,k}}}}$. In view of (\ref{eq:13}), (\ref{eq:53}) and employing Young's inequality, it is obtained that
\begin{flalign}
\dot{V}_{k}
\le& -\delta_1\underline{\lambda}
\left\|y\right\|^2-\sum\limits_{i = 1}^N\sum\limits_{q = 1}^{k} \jmath_{i,q}z_{i,q}^2- \sum\limits_{i = 1}^N\sum\limits_{q =1}^{k}\xi_{i,{q+1}}^{*}
\Theta_{i,{q+1}}^2&\nonumber\\
&+ \sum\limits_{i = 1}^N \sum\limits_{q = 1}^{k-1} {\tilde \Phi_{i,q}^T {{S_{i,q}}\left( {{\beta _{i,q}}} \right){z_{i,q}} -\sum\limits_{i = 1}^N\sum\limits_{q = 1}^{k-1} \tilde \Phi_{i,q}^T\Lambda _{i,q}^{-1}{{\dot {\hat \Phi}}_{i,q}}}}&\nonumber\\
&+ \sum\limits_{i = 1}^N {{z_{i,k}}{z_{i,k + 1}}}+\sum\limits_{i = 1}^N\sum\limits_{q =1}^{k}\varsigma_{i,q}^2+\Sigma_{k}&\label{eq:54}
\end{flalign}
where the fact that $-\left(\gamma_{i,k}+\frac{\dot{g}_{i,k}}
{2g_{i,k}^2\eta_i}\right)\le -\left(\gamma_{i,k}-\frac{\bar{g}_{i,d}}
{2\underline{g}_{i,k}^2\underline{\eta}_i}\right)$ is exploited, $\jmath_{i,k}= \gamma_{i,k}-\frac{\bar{g}_{i,d}}
{2\underline{g}_{i,k}^2\underline{\eta}_i}>0$ by choosing $\gamma_{i,k}$ large enough, $\xi_{i,k+1}^*>0$ is the design parameter, satisfying $\frac{1}{\xi_{i,k+1}}\ge \xi_{i,k+1}^*+ \frac{3}{4}$, ${\varsigma_{i,k}}=- {\dot \alpha _{i,k}}$, and $\Sigma_{k}=\Sigma_{k-1}+\sum\nolimits_{i = 1}^N\frac{1}{2}\bar{\epsilon}_{i,km}^2$.

$\textbf{Step \emph{n}}$:
Define a Lyapunov function ${V_n} = {V_{n - 1}} + \sum\nolimits_{i = 1}^N\frac{1}{2g_{i,n}\eta_i}z_{i,n}^2 + \sum\nolimits_{i = 1}^N \frac{1}{2}\tilde \Phi_{i,n}^T\Lambda_{i,n}^{ - 1}{{\tilde \Phi}_{i,n}}$. Taking the derivative of $V_n$ and using (\ref{eq:6}), (\ref{eq:1a}) and (\ref{eq:8}), yields
\begin{flalign}
\dot{V}_{n} \le& -\delta_1\underline{\lambda}
\left\|y\right\|^2-\sum\limits_{i = 1}^N\sum\limits_{q = 1}^{n-1} \jmath_{i,q}z_{i,q}^2-\sum\limits_{i = 1}^N\frac{\dot{g}_{i,n}}{2g_{i,n}^2
\eta_i}z_{i,n}^2&\nonumber\\
&-\sum\limits_{i =1}^N\frac{\dot{\eta}_{i}}{2g_{i,n}\eta_i^2}z_{i,n}^2+ \sum\limits_{i = 1}^N \sum\limits_{q = 1}^{n - 1} \tilde \Phi_{i,q}^T {S_{i,q}}\left( {{\beta _{i,q}}} \right){z_{i,q}}&\nonumber\\
&- \sum\limits_{i = 1}^N \sum\limits_{q = 1}^{n - 1} \tilde \Phi_{i,q}^T\Lambda_{i,q}^{ - 1}{{\dot {\hat \Phi}}_{i,q}}- \sum\limits_{i = 1}^N\sum\limits_{q=1}^{n-1}\xi_{i,{q+1}}^{*}
\Theta_{i,{q+1}}^2&\nonumber\\
&+ \sum\limits_{i = 1}^N{z_{i,n}}{u_i} + \sum\limits_{i = 1}^N {{z_{i,n}}\left({z_{i,{n-1}}} + {\Psi_{i,n}}\left( {{\beta _{i,n}}} \right)\right)}& \nonumber\\
&-\sum\limits_{i = 1}^N\tilde \Phi_{i,n}^T\Lambda_{i,n}^{ - 1}{{\dot{\hat \Phi}}_{i,n}}+\sum\limits_{i =1}^N\sum\limits_{q=1}^{n-1}
\varsigma_{i,q}^2 +\Sigma_{n-1}.&\label{eq:55}
\end{flalign}
where ${\Psi_{i,n}}\left( {{\beta _{i,n}}} \right)= \frac{1}{g_{i,n}\eta_i} ({\dot{\eta}}_i
{x_{i,n}}+{\eta}_if_{i,n}\left(\check{x}_{i,n},0\right)+
{\eta}_i d_{i,n}-{\dot{\alpha} _{i,nf}})$ is   approximated by utilizing the RBFNN, with ${{\beta_{i,n}}}=[\check{x}_{i,n},{x}_{j,1},
\eta_i,\dot{\eta}_i]^{T} \in{\Omega_{{\beta _{i,n}}}}$.
The distributed controller $v_i$ in (\ref{eq:48}) can be written in the form of
\begin{flalign}
{v_i} = & -{\gamma_{i,n}}{z_{i,n}}-{z_{i,n}}- {z_{i,n - 1}} - \hat \Phi_{i,n}^T{S_{i,n}}\left( {{\beta _{i,n}}} \right)& \nonumber\\
& + \gamma_{i,n}\left( {{z_{i,n}} - {{\bar z}_{i,n}}} \right)+\left( {{z_{i,n}} - {{\bar z}_{i,n}}} \right)+ {z_{i,n - 1}}& \nonumber\\
&- {{\bar z}_{i,n - 1}}+ \hat \Phi_{i,n}^T \left({S_ {i,n}}\left( {{\beta _{i,n}}} \right) - {S_{i,n}}\left( {{{\bar \beta }_{i,n}}} \right)\right). &\label{eq:56}
\end{flalign}
Inserting (\ref{eq:49}) and (\ref{eq:56}) into (\ref{eq:55}), ${{\dot V}_n}$ becomes
\begin{flalign}
{{\dot V}_n} \le&  -\delta_1\underline{\lambda}\left\|y\right\|^2- \sum\limits_{i = 1}^N {\sum\limits_{q = 1}^{n}{{\jmath_{i,q}}z_{i,q}^2}}- \sum\limits_{i = 1}^N {\sum\limits_{q = 1}^{n - 1} {\xi_{i,q + 1}^*\Theta_{i,q + 1}^2}}&\nonumber\\
&+\sum\limits_{i = 1}^N \sum\limits_{q = 1}^N\tilde \Phi_{i,q}^T \left( {{S_{i,q}}\left( {{\beta _{i,q}}} \right){z_{i,q}} - {S_{i,q}}\left( {{{\bar \beta }_{i,q}}} \right){{\bar z}_{i,q}}} \right)&\nonumber\\
&+\sum\limits_{i = 1}^N\left|z_{i,n}\right|
\left({\Delta{\alpha}_{i,n}}+\Delta{u_i}\right) + \sum\limits_{i = 1}^N {\sum\limits_{q = 1}^{n - 1} {\varsigma_{i,q}^2} }+\bar{H} & \label{eq:58}
\end{flalign}
with
\begin{flalign}
\Delta {\alpha_{i,n}} = &\,{{\gamma_{i,n}}\Delta {z_{i,n}} + \Delta {z_{i,n}}} + \Delta {z_{i,n - 1}}&\nonumber\\
&\,+\left| {\hat \Phi_{i,n}^T {{S_{i,n}}\left( {{\beta _{i,n}}} \right) - \Phi_{i,n}^T{S_{i,n}}
\left( {{{\bar \beta }_{i,n}}} \right)}} \right|& \label{eq:58a}
\end{flalign}
where the fact that $-\left(\gamma_{i,n}+\frac{\dot{g}_{i,n}}
{2g_{i,n}^2\eta_i}+\frac{\dot{\eta}_{i}}{2g_{i,n}
\eta_i^2}\right)\le -\left(\gamma_{i,n}-\frac{\bar{g}_{i,d}}
{2\underline{g}_{i,n}^2\underline{\eta}_i}
-\frac{\bar{\eta}_{i,d}}{2\underline{g}_{i,n}
\underline{\eta}_i^2}\right)$ is utilized, $\jmath_{i,n}=\gamma_{i,n}-\frac{\bar{g}_{i,d}}{2\underline{g}_{i,n}^2
\underline{\eta}_i}-\frac{\bar{\eta}_{i,d}}
{2\underline{g}_{i,n}
\underline{\eta}_i^2}>0$ by choosing $\gamma_{i,n}$ large enough, and $\bar{H}=\sum\nolimits_{i =1}^N\frac{1}{4}\bar{\epsilon}_{i,nm}^2+\Sigma_{n-1}$.

${\textbf{Remark 6}}$.
Here, we pause to stress that several triggering error terms that affect the stability of the system emerge due to the participation of event-triggering mechanism, that is, $\left|z_{i,n}\right|{\Delta{\alpha}_{i,n}}$, $\left|z_{i,n}\right|\Delta{u_i}$ and ${\tilde \Phi_{i,q}^T\left( {{S_{i,q}}\left( {{\beta _{i,q}}} \right){z_{i,q}} -{S _{i,q}}\left( {{{\bar \beta }_{i,q}}} \right){{\bar z}_{i,q}}} \right)}$, $q=1,\cdots,n$. How to tackle such adverse effects to ensure the stability of the system is one of the main challenges in achieving the control objectives of this paper.
In the following, we will show that they are bounded by vital results exhibited in \emph{Lemmas} 2-3, and can be incorporated into $\dot{V}_n$. The impacts of triggering error, as a result, is handled.

By employing \emph{Lemmas} 2-3, we can obtain
\begin{flalign}
&\left|z_{i,n}\right|({\Delta {\alpha}_{i,n}}+\Delta u_i)\le \frac{1}{2}z_{i,n}^2+{\Delta \alpha_{i,n}^2 }+{\Delta u_i^2 }&\label{eq:59}\\
&\sum\limits_{i = 1}^N \sum\limits_{q = 1}^N\tilde \Phi_{i,q}^T \left( {{S_{i,q}}\left( {{\beta _{i,q}}} \right){z_{i,q}} - {S_{i,q}}\left( {{{\bar \beta }_{i,q}}} \right){{\bar z}_{i,q}}} \right)&\nonumber\\
&\le\,\sum\limits_{i = 1}^N\sum\limits_{q = 1}^n {\frac{1}{4}z_{i,q}^2}  + \sum\limits_{i = 1}^N\sum\limits_{q = 1}^n {{\Delta s_{i,q}^2}} \tilde \Phi_{i,q}^T{{\tilde \Phi}_{i,q}} & \nonumber\\
&+\sum\limits_{i = 1}^N\sum\limits_{q = 1}^n\frac{\sigma_{i,w}}{4}\tilde \Phi_{i,q}^T{{\tilde \Phi}_{i,q}}+ \sum\limits_{i = 1}^N\sum\limits_{q = 1}^n \frac{\bar{s}_{i,q}^2\Delta z_{i,q}^2}{\sigma_{i,w}} & \label{eq:61}
\end{flalign}
where $\sigma_{i,w}>0$ is the design parameter. According to (\ref{eq:59}), (\ref{eq:61}) and exploits the fact that $\left\|\tilde \Phi_{i,k}\right\|\le \phi_{i,km}$, $k=1,\cdots,n$, it is seen that $\dot{V}_{n}$ in (\ref{eq:58}) becomes
\begin{flalign}
{{\dot V}_n}\le &  -\delta_1\underline{\eta}\left\|y\right\|^2- \sum\limits_{i = 1}^N {\sum\limits_{q = 1}^n {\jmath_{i,q}^{*}z_{i,q}^2}} +\sum\limits_{i = 1}^N {\sum\limits_{q = 1}^{n - 1} {\varsigma_{i,q}^2} } &\nonumber\\
&- \sum\limits_{i = 1}^N \sum\limits_{q = 1}^n \frac{\sigma_{i,w}}{2}\tilde \Phi_{i,q}^T{{\tilde \Phi}_{i,q}} - \sum\limits_{i = 1}^N {\sum\limits_{q = 1}^{n - 1} {\xi_{i,q + 1}^{*}\Theta_{i,q + 1}^2}}& \nonumber\\
&+\sum\limits_{i = 1}^N \sum\limits_{q = 1}^n \left(\frac{{\Delta s_{i,q}^2}} {{{\gamma_{i,q}}}}+\frac{\sigma_{i,w}}{4} \right)\phi_{i,qm}^2+{\Xi_n}& \label{eq:a63}
\end{flalign}
where $\jmath_{i,q}^*=\jmath_{i,q}-\frac {1}{4}>0$, $\jmath_{i,n}^*=\jmath_{i,n}-\frac {3}{4}>0$ by choosing $\gamma_{i,k}$ large enough, for $q=1,\cdots,n-1$, $k=1,\cdots,n$, ${\Xi_n} =\bar{H}+\sum\nolimits_{i = 1}^N{\Delta \alpha_{i,n}^2 +\sum\nolimits_{i = 1}^N\Delta u_i^2}
+\sum\nolimits_{i = 1}^N\sum\nolimits_{q = 1}^n \frac{\bar{s}_{i,q}^2\Delta z_{i,q}^2}{\sigma_{i,w}}$.

Since the set ${\Omega _{v}}:=\{\sum\nolimits_{i = 1}^N {\sum\nolimits_{q =1}^{n-1}}\frac{1}
{g_{i,q}}{z_{i,q}^2}+\sum\nolimits_{i = 1}^N \frac{1}{g_{i,n}\eta_i}{z_{i,n}^2}
+\sum\nolimits_{i = 1}^N {\sum\nolimits_{q = 1}^{n}} {\tilde \Phi_{i,q}^T\Lambda_{i,q}^{ - 1}{{\tilde \Phi}_{i,q}}}+ \sum\nolimits_{i = 1}^N {\sum\nolimits_{q = 1}^{n-1}}{\Theta_{i,q + 1}^2}\le 2\varrho_0\}$ is a compact one for any $\varrho_0>0$, which implies that, for all $q=1,\cdots,n-1$, $\left| {{\varsigma_{i,q}}} \right| \le \bar{\varsigma}_{i,q}$ on $\Omega_v$. This, together with (\ref{eq:a63}), results in
\begin{flalign}
{{\dot V}_n}\le & - \sum\limits_{i = 1}^N {\sum\limits_{q = 1}^n{\jmath_{i,q}^{*}z_{i,q}^2}}- \sum\limits_{i = 1}^N \sum\limits_{q = 1}^n \frac{\sigma_{i,w}}{2}\tilde \Phi_{i,q}^T{{\tilde \Phi}_{i,q}}\nonumber\\
& - \sum\limits_{i = 1}^N {\sum\limits_{q = 1}^{n-1}{\xi_{i,q + 1}^*\Theta_{i,q + 1}^2}} -\delta_1\underline{\eta}\left\|y\right\|^2+ {\Sigma_n} & \nonumber\\
\le& -\delta_1\underline{\eta}\left\|y\right\|^2
-\hbar_1{V_n}+{\Sigma_n}& \label{eq:63}
\end{flalign}
where $\hbar_1=\min\{2{\bar{g}_{i,1}
\jmath_{i,1}^{*}},\cdots,2{\bar{g}_{i,n}\jmath_{i,n}^{*}}$,
$\frac{{\sigma_{i,1}}}{{\lambda_{\max}}\left\{ {\Lambda_{i,1}^{-1}} \right\}},\cdots,\frac{{\sigma _{i,n}}}{{\lambda_{\max }}\left\{{\Lambda_{i,n}^{-1}} \right\}}, 2\xi_{i,2}^*,\cdots,2\xi_{i,n}^*\}$ and
${\Sigma_n}={\Xi_n}+\sum\nolimits_{i = 1}^N \sum\nolimits_{q = 1}^n \left(\frac{{\Delta s_{i,q}^2}}{{{\gamma_{i,q}}}}+\frac{\sigma_{i,w}}{4} \right)\phi_{i,qm}^2+ \sum\nolimits_{i = 1}^N {\sum\nolimits_{q = 1}^{n-1}} {\bar{\varsigma}_{i,q}^2}$.

Next, we show that the results in \emph{Theorem} 2 hold.
\begin{itemize}
\item [i)]{In view of (\ref{eq:63}), it is seen that ${V_{n}} \in {L_\infty}$, it indicates that ${z_{i,k}}$, ${\tilde \Phi_{i,k}}$ and ${\Theta_{i,q + 1}}$ are all bounded, $k = 1, \cdots ,n$, $q = 1, \cdots ,n - 1$. This, along with (\ref{eq:7})-(\ref{eq:9}) and  (\ref{eq:11})-(\ref{eq:13}), results in $x_{i,k}\in {L_\infty }$. It can be further deduced from (\ref{eq:48}) that $v_i\in {L_\infty }$.}
\item [ii)]{Recalling (\ref{eq:7}) and the definitions of $V_1$ and $V_n$, one can obtain that ${\left\|y\right\|^2}= \sum\nolimits_{i = 1}^N{\frac{1}{\eta_i^2}z_{i,1}^2} \le b{V_n}$, which implies that ${\left\|y \right\|^2} \le b\left[ {{e^{ - \hbar_1 t}}{V_n}\left( 0 \right) + \frac{{{\Sigma_n}}}{\hbar_1}\left( {1 - {e^{ - \hbar_1 t}}} \right)} \right]$, where $b=\max\left\{1,\frac{2\bar{g}_{1,1}}
    {\underline{\eta}_1^2},\cdots,\frac{2\bar{g}_{N,1}}
    {\underline{\eta}_N^2}\right\}$ is some positive constant. Thus we can conclude that the outputs of all the subsystems reach a consensus against sensor failures. In addition, it is readily seen from (\ref{eq:63}) that $\|y(t)\|_{[0, T]}\leq \sqrt{\frac{1}{\delta_1\underline{\lambda} }\left[\frac{V_{n}(0)}{T}+\Sigma_n\right]}$,
    it indicates that the upper bound for ${\left\|y\right\|_{[0, T]}}$ can be decreased by decreasing the triggering thresholds $\Delta x_{i,k}$, $\Delta x_{j,k}$ and $\Delta u_i$, and increasing design parameters $\gamma_{i,k}$, $\delta_1$, $\Lambda_{i,k}$ and $\xi_{i,q}$, $k=1,\cdots,n$, $q=2,\cdots,n$.}
\item [iii)]{For all $t \in\left[t_{k,p}^{i}, t_{k,p+1}^{i} \right)$, $k=1,\cdots,n$, $p=0,1,2,\cdots$, we define $\omega_{k,p}^{i}(t)=x_{i,k}(t) -\bar{x}_{i,k}(t)$, then it follows that
    \begin{flalign}
    \frac{d\left|\omega_{k,p}^{i}\right|}{d t}=&\, \frac{d\left(\omega_{k,p}^{i} {\rm{\times}} \omega_{k,p}^{i}\right)^{\frac{1}{2}}}{d t}&\nonumber\\
    =&\, {\rm{sign}}\left(\omega_{k,p}^{i}\right) \dot{\omega}_{k,p}^{i} \leq
    \left|\dot{\omega}_{k,p}^{i}\right|.
    &\label{eq:65}
    \end{flalign}
    Since $\bar{x}_{i,k}(t)$ remains unchanged for $t \in\left[t_{k,p}^{i}, t_{k,p+1}^{i}\right)$, it is seen that, for $k=1,\cdots,n$
    \begin{flalign}
    \left|\dot{\omega}_{k,p}^{i} \right|=&\,{g_{i,k}}\left(\check{x}_{i,k+1}\right){x_{i,{k+1}}}
    +{f_{i,k}}\left({{\check{x}_{i,{k}}},0}\right)&\nonumber\\
    &+d_{i,k}\left(\check{x}_{i,k+1}\right)&\label{eq:a65}
    \end{flalign}
    As $x_{i,k+1}$, $g_{i,k}$, $f_{i,k}$ and $d_{i,k}$ in (\ref{eq:a65}) are all bounded, it holds that $\left|\dot{\omega}_{k,p}^{i}\right| \leq x_{0}^{i}$, $k=1,\cdots,n$, where $x_{0}^{i}>0$ is an unknown constant, which implies that
    $t_{k,p+1}^{i}-t_{k,p}^{i} \geq \Delta x_{i,k}/ x_{0}^{i}=T_{x}^{i}>0$. By following the similar analysis, it is seen that   $t_{k,p+1}^{j}-t_{k,p}^{j}\geq \Delta x_{j,k}/ x_{0}^{j}=T_{x}^{j}>0$, $k=1,\cdots,n$, and $t_{u,p+1}^{i}-t_{u,p}^{i}\geq \Delta u_{i}/ u_{0}^{i}=T_{u}^{i}>0$, where $x_{0}^{j}>0$ and $u_{0}^{i}>0$ are some unknown constants. Thus the Zeno solutions are ruled out. $\hfill\blacksquare$}
\end{itemize}

${\textbf{Remark 7}}$.
Recalling the definitions of ${\Psi_{i,k}}\left( {{\beta _{i,k}}} \right)$, $k=1,\cdots,n$, it is deduced that ${\beta _{i,1}}$ is a function of variables ${x}_{i,1}$, $\eta_i$ and $\dot{\eta}_i$, and ${\beta _{i,k}}$ is a function of variables $\check{x}_{i,k}$, ${x}_{j,1}$, $\hat {\Phi}_{i,{1}},\cdots, \hat {\Phi}_{i,{k-1}}$, $\eta_i$ and $\dot{\eta}_i$, $k=2,\cdots,n$. However, it should be noted that regarding weights $\hat {\Phi}_{i,{1}},\cdots, \hat {\Phi}_{i,{k-1}}$ as NN inputs is inadvisable since the curse of dimensionality of RBFNN may result in a larger number of NN inputs. To reduce the NN input dimension and thus the computational cost, in this work we choose $\beta _{i,1}=[x_{i,1}, \eta_i,\dot{\eta}_i]^{T} \in{\Omega_{\beta_{i,1}}}$ and ${\beta _{i,k}}=\left[\check{x}_{i,k},{x}_{j,1},
\eta_i,\dot{\eta}_i\right]^{T}
\in{\Omega_{\beta_{i,k}}}$, $k=2,\cdots,n$.

${\textbf{Remark 8}}$.
The problem of solving non-differentiable virtual control signals for nonlinear systems has been reported in recent works  \cite{zhang2021adaptive,sun2022Distributed,zhang2021event} in the framework of state-triggering control, therein the repeated differentiation issue is avoided by employing the first-order filter. Nevertheless, the output of the filter is required to be triggered to derive the system stability, which inevitably increases the computational burden of the sensors.
Besides, these solutions concentrate on the case where  only the plant states or the control input are transmitted over the network, leaving sensor failures unaccounted for.
Moreover, the systems considered in \cite{zhang2021adaptive,sun2022Distributed,zhang2021event} are limited to strict-feedback form. This paper develops a distributed adaptive control scheme under  event-triggering setting for more general uncertain non-affine pure-feedback nonlinear MASs with sensor failures, in which the restriction on triggering the filter-output is lifted by applying the projection operator in the parameter update law (\ref{eq:49}).
Additionally, the sensor failures are effectively handled without using fault detection and diagnosis unit or controller reconfiguration, which is contrary to the related works  \cite{zhang2021event,zhai2018output,zhang2018observer}.

\begin{figure}
\begin{center}
\includegraphics[width=0.34\textwidth,height=44mm]{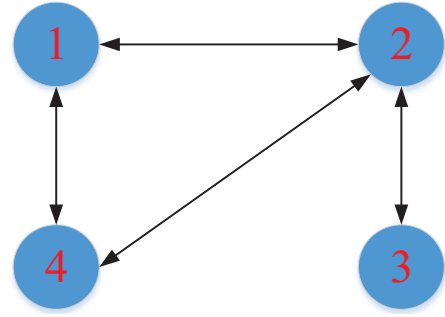}
\caption{The communication graph.}
\end{center}
\end{figure}

\section{Simulation Studies}
Consider a MAS comprised of 4 nonlinear subsystems, with the $i$th ($i=1,\cdots,4$) agent taking the form of:
\begin{flalign}
{{\dot x}_{i,1}}=\,&{x_{i,1}}+{x_{i,2}}
+\frac{1}{5}{x_{i,2}^3}& \nonumber\\
{{\dot x}_{i,2}}=\,&{x_{i,1}}{x_{i,2}}+{u_{i}} +\frac{1}{7}{u_{i}^3}& \nonumber\\
{y_i} =\,& {x_{i,1}}& \label{eq:66}
\end{flalign}
for $i=1,\cdots,4$, $k=1,2$.
The communication topology is presented in Fig. 2. The initial values $\check{x}_1(0)=\left[0.1, 0.05, 0.1, 0.05\right]^T$, $\check{x}_{2}(0)=\left[0.1, 0.1, 0.1, 0.1\right]^T$, the fault factor $\eta_i=0.6$, $\tau_f=1$, which indicates that the sensor loses $60\%$ of its effectiveness at $t= 1s$, the design parameters $\delta_1=0.5$, $\gamma_{i,1}=5$, $\gamma_{i,2}=10$, $\sigma_{i,1}=20$, $\sigma_{i,2}=20$, $\Lambda_{i,1}=0.005$, $\Lambda_{i,2}=0.005$ and ${\xi_{i,2}}=0.01$, the triggering thresholds $\Delta x_{i,k}=0.001$, $\Delta x_{j,k}=0.001$ and $\Delta u_{i}=0.01$. The NN contains 25 nodes, and $\zeta_b=2$. Figs. 2-3 display the output trajectories of $x_{i,1}^f$ and $x_{i,2}^f$, respectively, from which it can be observed that the outputs of all the subsystems reach a consensus in the presence of sensor failures. From Fig. 4, we can see that consensus error $e_i$ converges into residual set near the origin. The boundedness of distributed control $u_i$ is depicted in Fig. 5.

To test the impact of triggering thresholds on the performance of the system, another set of triggering thresholds $\Delta x_{i,k}^{\prime}=0.002$, $\Delta x_{j,k}^{\prime}=0.005$ and $\Delta u_{i}^{\prime}=0.05$ are chosen for comparison. 
The evolution of the consensus error $e_i$ is shown in Figs. 6. The triggered times of $x_{i,1}^f$, $x_{i,2}^f$ and $u_i$ under different triggering thresholds are presented in Figs. 7-9, respectively.
From which we can conclude that the triggering times for communication decrease as the triggering thresholds increase, leading to savings in communication resources. However, increasing triggering thresholds also leads to an increase of the consensus error $e_i$.

\begin{figure}
\begin{center}
\includegraphics[width=0.45\textwidth,height=53mm]{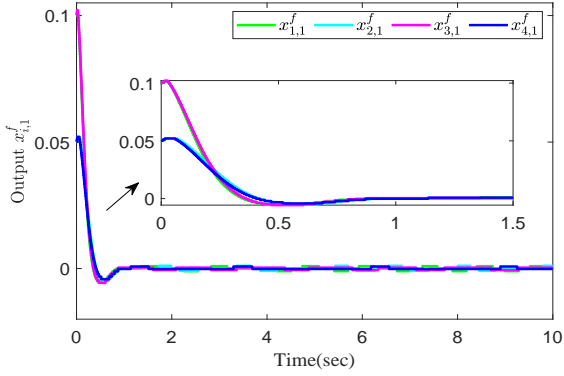}
\caption{The output $x_{i,1}^f$.}
\end{center}
\end{figure}

\begin{figure}
\begin{center}
\includegraphics[width=0.45\textwidth,height=53mm]{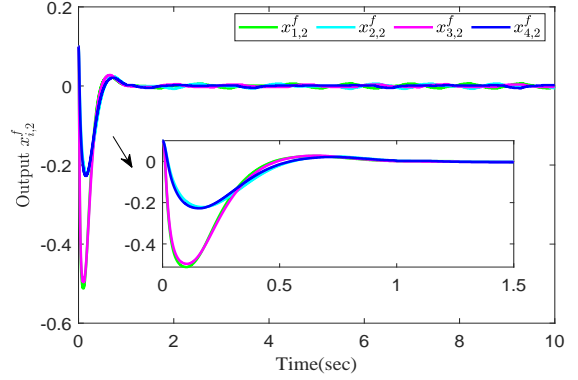}
\caption{The output $x_{i,2}^f$.}
\end{center}
\end{figure}

\begin{figure}
\begin{center}
\includegraphics[width=0.45\textwidth,height=53mm]{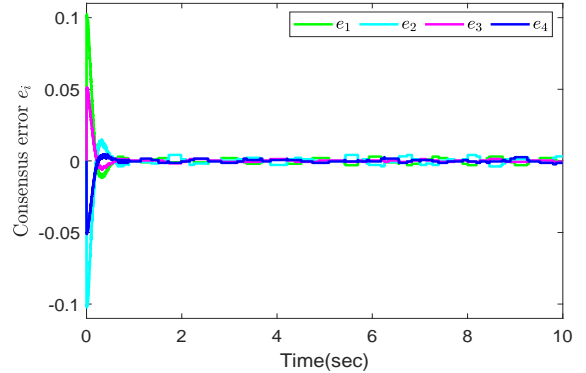}
\caption{Consensus error $e_i$.}
\end{center}
\end{figure}

\begin{figure}
\begin{center}
\includegraphics[width=0.45\textwidth,height=53mm]{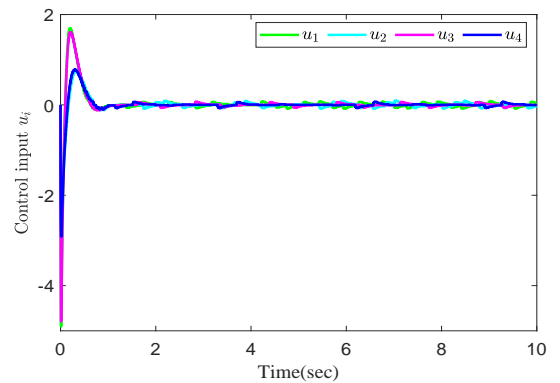}
\caption{Distributed controller $u_i$.}
\end{center}
\end{figure}

\begin{figure}
\begin{center}
\includegraphics[width=0.45\textwidth,height=53mm]{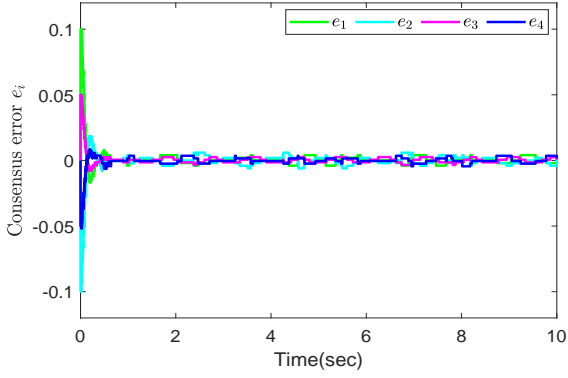}
\caption{$e_i$ in the case of increasing triggering thresholds.}
\end{center}
\end{figure}

\begin{figure}
\begin{center}
\includegraphics[width=0.45\textwidth,height=53mm]{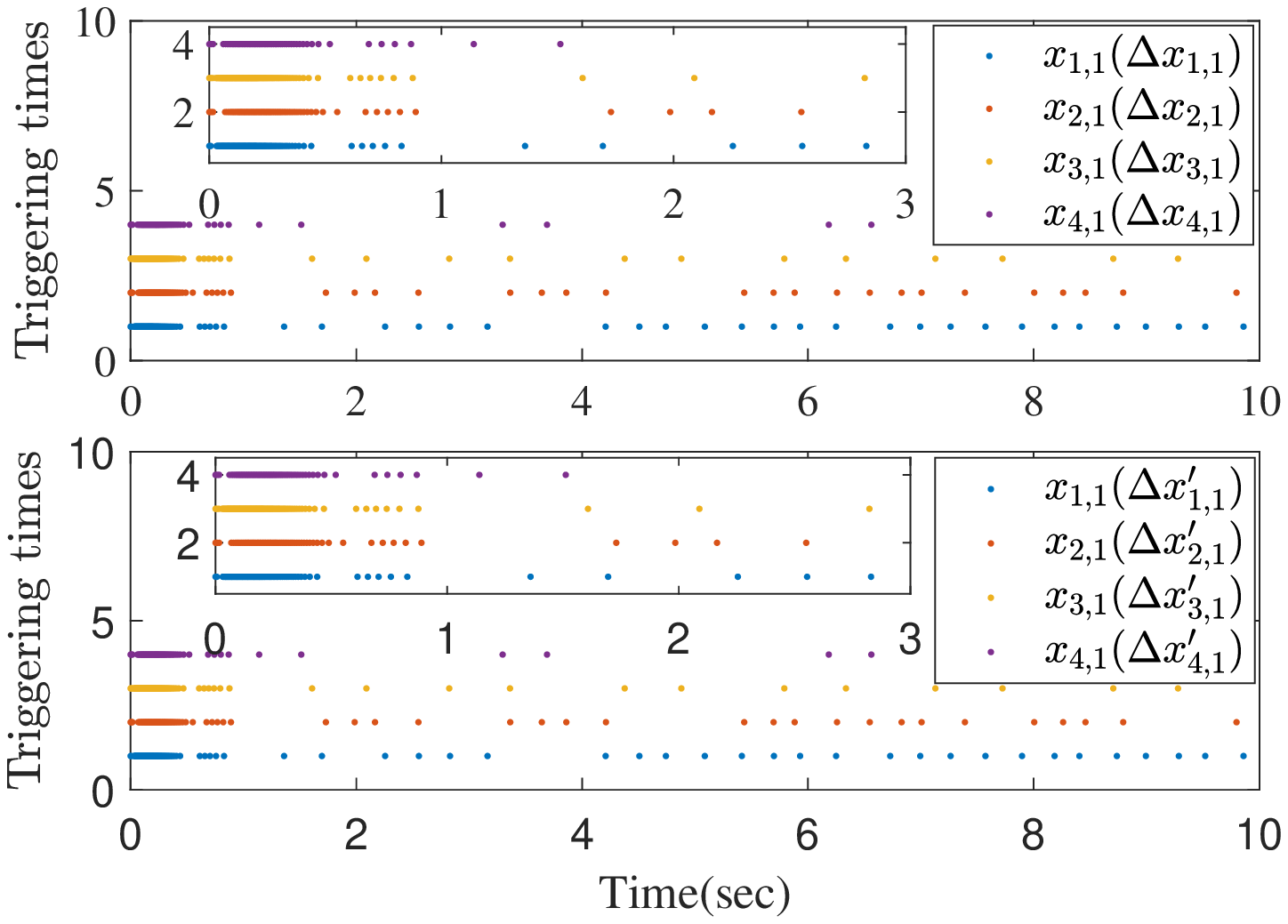}
\caption{Triggering times of $x_{i,1}$ for different triggering thresholds.}
\end{center}
\end{figure}

\begin{figure}
\begin{center}
\includegraphics[width=0.45\textwidth,height=53mm]{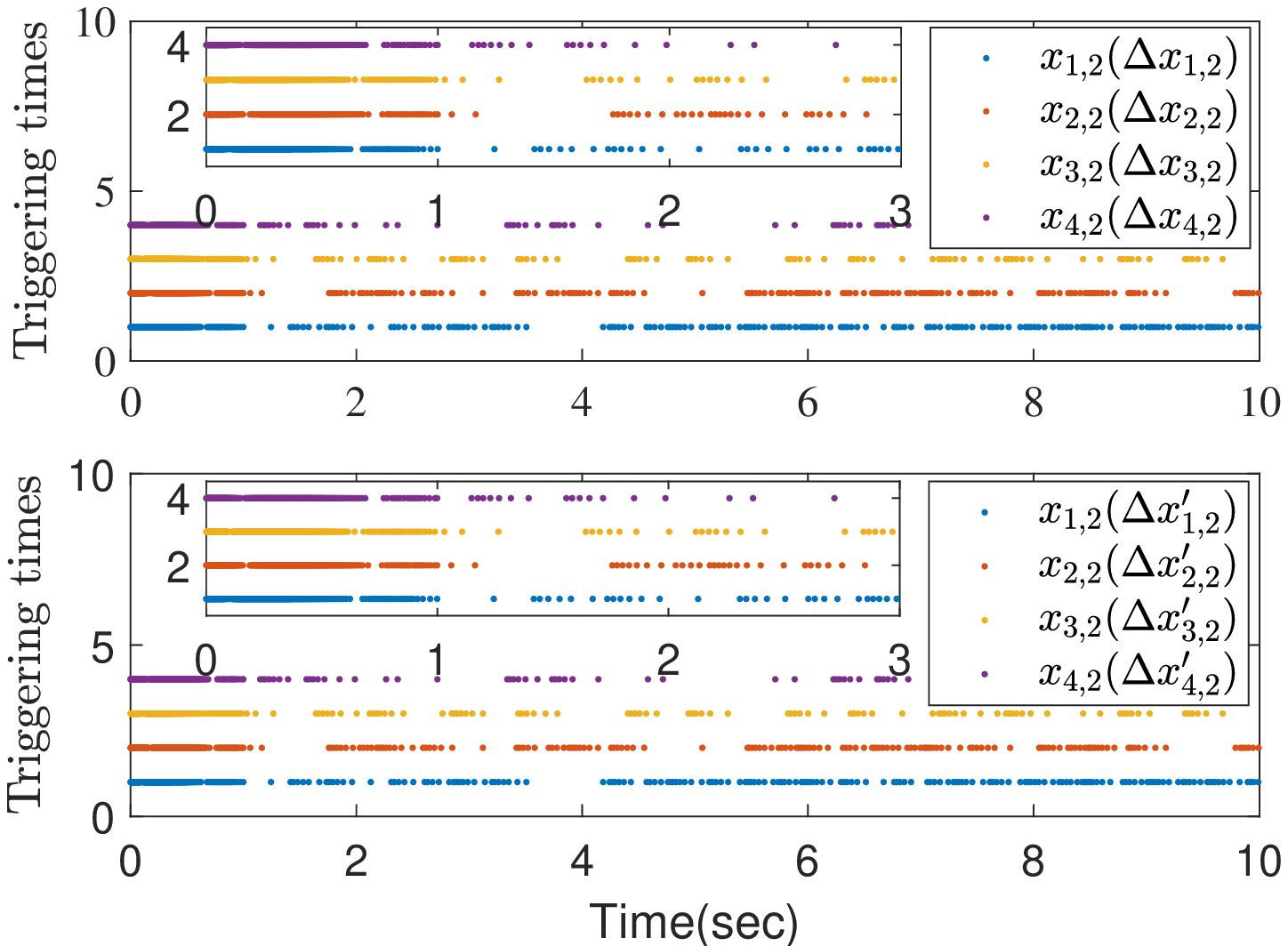}
\caption{Triggering times of $x_{i,2}$ for different triggering thresholds.}
\end{center}
\end{figure}

\begin{figure}
\begin{center}
\includegraphics[width=0.45\textwidth,height=53mm]{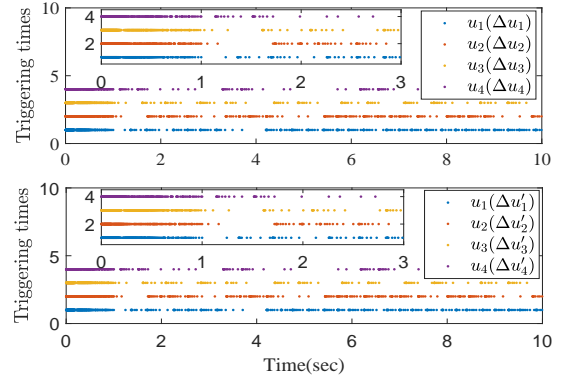}
\caption{Triggering times of $u_{i}$ for different triggering thresholds.}
\end{center}
\end{figure}

\section{Conclusion}
This paper develops a distributed adaptive control algorithm for uncertain non-affine pure-feedback nonlinear MASs under event-triggering setting, wherein all state variables are not directly and continuously available or even polluted due to sensor failures. By fusing a novel replacement policy, NN approximating and dynamic filtering technique, it is shown that all the internal signals are SGUUB, with the outputs of all the subsystems reaching a consensus, while precluding infinitely fast execution. An attractive topic of future research is to consider the tracking control problem for such system.

\begin{appendices}
\section{}
$\textbf{Step 1}$: Define ${V_1}= \sum\nolimits_{i = 1}^N\frac{1}{2g_{i,1}}z_{i,1}^2 + \sum\nolimits_{i = 1}^N \frac{1}{2}\tilde \Phi_{i,1}^T\Lambda_{i,1}^{ - 1}{{\tilde \Phi}_{i,1}} + \sum\nolimits_{i = 1}^N \frac{1}{2}\Theta_{i,2}^2$. In accordance with (\ref{eq:6}), (\ref{eq:1a}), (\ref{eq:7}), (\ref{eq:8}), (\ref{eq:11}), (\ref{eq:12}) and (\ref{eq:15}), the time derivative of $V_1$ is obtained as
\begin{flalign}
\dot{V}_{1} \le & -\delta_1\underline{\lambda}
\left\|y\right\|^2-\sum\limits_{i = 1}^N \jmath_{i,1}z_{i,1}^2-\sum\limits_{i = 1}^N\frac{\sigma_{i,1}}{2}\tilde \Phi_{i,1}^T{{{\tilde \Phi}}_{i,1}}&\nonumber\\
& - \sum\limits_{i = 1}^N\xi_{i,2}^{*}\Theta_{i,2}^2 +\sum\limits_{i = 1}^N{z}_{i,1}z_{i,2}+\sum\limits_{i = 1}^N\varsigma_{i,1}^2+\Sigma_1&\label{eq:25}
\end{flalign}
where $\jmath_{i,1}=\gamma_{i,1} -\frac{\bar{g}_{i,d}}{2\underline{g}_{i,1}^2}>0$ by choosing $\gamma_{i,1}$ large enough, $\xi_{i,2}^*>0$ is the design parameter, satisfying $\frac{1}{\xi_{i,2}}\ge \xi_{i,2}^*+ \frac{3}{4}$, ${\varsigma_{i,1}}=-{\dot \alpha _{i,1}}$, ${\Sigma_1} =\sum\nolimits_{i = 1}^N{\frac{{{\sigma _{i,1}}}}{2}{\phi_{i,1m}^2}}  + \sum\nolimits_{i = 1}^N\frac{1}{2}\bar{\epsilon}_{i,1m}^2$, and the fact that $-\left(\gamma_{i,1}+\frac{\dot{g}_{i,1}}
{2{g}_{i,1}^2}\right)\le -\left(\gamma_{i,1}
-\frac{\bar{g}_{i,d}}{2\underline{g}_{i,1}^2}
\right)$ is exploited.

$\textbf{Step \emph{k}}$ $(k=2,\cdots,n-1)$:
Define ${V_k} = {V_{k - 1}} + \sum\nolimits_{i = 1}^N \frac{1}{2g_{i,k}}z_{i,k}^2 + \sum\nolimits_{i = 1}^N \frac{1}{2}\tilde \Phi_{i,k}^T\Lambda_{i,k}^{-1}{{\tilde \Phi}_{i,k}}+ \sum\nolimits_{i = 1}^N \frac{1}{2}\Theta_{i,k + 1}^2$. In view of (\ref{eq:6}), (\ref{eq:1a}), (\ref{eq:8}), (\ref{eq:11}), (\ref{eq:13}),  (\ref{eq:15}) and (\ref{eq:25}), we can obtain
\begin{flalign}
\dot{V}_{k}\le& -\delta_1\underline{\lambda}\left\|y\right\|^2
-\sum\limits_{i = 1}^N\sum\limits_{q = 1}^{k} \jmath_{i,q}z_{i,q}^2+ \sum\limits_{i = 1}^N {{z_{i,k}}{z_{i,k + 1}}}&\nonumber\\
&-\sum\limits_{i = 1}^N\sum\limits_{q =1}^{k}\frac{\sigma_{i,q}}{2}\tilde \Phi_{i,q}^T{{{\tilde \Phi}}_{i,q}}- \sum\limits_{i = 1}^N\sum\limits_{q =1}^{k}\xi_{i,{q+1}}^{*}
\Theta_{i,{q+1}}^2&\nonumber\\
&+\sum\limits_{i = 1}^N\sum\limits_{q =1}^{k}\varsigma_{i,q}^2+\Sigma_{k}&\label{eq:30}
\end{flalign}
where $\jmath_{i,k}=\gamma_{i,k}-\frac{\bar{g}_{i,d}}
{2\underline{g}_{i,k}^2\underline{\eta}_i}>0$ by choosing $\gamma_{i,k}$ large enough, $\xi_{i,k+1}^*>0$ is the design parameter, satisfying  $\frac{1}{\xi_{i,k+1}}\ge \xi_{i,k+1}^*+ \frac{3}{4}$, ${\varsigma_{i,k}}=- {\dot \alpha _{i,k}}$, $\Sigma_{k}=\Sigma_{k-1}+\sum\nolimits_{i = 1}^N{\frac{{{\sigma _{i,k}}}}{2}{{\phi_{i,km}^2}}}+\sum\nolimits_{i = 1}^N\frac{1}{2}\bar{\epsilon}_{i,km}^2$, and the fact that $-\left(\gamma_{i,k}+\frac{\dot{g}_{i,k}}
{2g_{i,k}^2\eta_i}\right)\le -\left(\gamma_{i,k}-\frac{\bar{g}_{i,d}}
{2\underline{g}_{i,k}^2\underline{\eta}_i}\right)$ is used.

$\textbf{Step \emph{n}}$:
Define ${V_n} = {V_{n-1}} + \sum\limits_{i = 1}^N \frac{1}{2g_{i,n}\eta_i}z_{i,n}^2 + \sum\limits_{i = 1}^N \frac{1}{2}\tilde \Phi_{i,n}^T\Lambda_{i,n}^{ - 1}{{\tilde \Phi}_{i,n}}$.
Applying (\ref{eq:6}), (\ref{eq:1a}), (\ref{eq:8}), (\ref{eq:14}), (\ref{eq:15}) and (\ref{eq:30}), one can obtain
\begin{flalign}
\dot{V}_{n}\le&-\sum\limits_{i = 1}^N\sum\limits_{q = 1}^{n} \jmath_{i,q}z_{i,q}^2-\sum\limits_{i = 1}^N\sum\limits_{q =1}^{n}\frac{\sigma_{i,q}}{2}
\tilde \Phi_{i,q}^T{{{\tilde \Phi}}_{i,q}}&\nonumber\\
&- \sum\limits_{i = 1}^N\sum\limits_{q =1}^{n-1}\xi_{i,{q+1}}^{*}
\Theta_{i,{q+1}}^2 -\delta_1\underline{\lambda}
\left\|y\right\|^2+\Sigma_n& \nonumber\\
\le& -\delta_1\underline{\lambda}
\left\|y\right\|^2-\hbar_1 V_n+\Sigma_n& \label{eq:36}
\end{flalign}
where $\jmath_{i,n}= \gamma_{i,n}-\frac{\bar{g}_{i,d}}{2\underline{g}_{i,n}^2
\underline{\eta}_i}-\frac{\bar{\eta}_{i,d}}{2\underline{g}_{i,n}
\underline{\eta}_i^2}>0$ by choosing $\gamma_{i,n}$ large enough, $\hbar_1  =\min\{2{\bar{g}_{i,1}\jmath_{i,1}}
,\cdots,2{\bar{g}_{i,n}\jmath_{i,n}}$,
$\frac{{\sigma_{i,1}}}{{\lambda_{\max}}\left\{ {\Lambda_{i,1}^{-1}} \right\}},\cdots,\frac{{\sigma _{i,n}}}{{\lambda_{\max }}\left\{{\Lambda_{i,n}^{-1}} \right\}}, 2\xi_{i,2}^*,\cdots,2\xi_{i,n}^*\}$,  ${\Sigma_n}=\Sigma_{n-1}+\sum\nolimits_{i = 1}^N\frac{1}{4}
\bar{\epsilon}_{i,nm}^2+\sum\nolimits_{i = 1}^N {\frac{{{\sigma _{i,n}}}}{2}{\phi_{i,nm}^2}}  + \sum\nolimits_{i = 1}^N {\sum\nolimits_{q = 1}^{n-1}} {\bar{\varsigma}_{i,q}^2}$, and the fact that $-\left(\gamma_{i,n}+\frac{\dot{g}_{i,n}}
{2g_{i,n}^2\eta_i}+\frac{\dot{\eta}_{i}}{2g_{i,n}
\eta_i^2}\right)\le -\left(\gamma_{i,n}-\frac{\bar{g}_{i,d}}
{2\underline{g}_{i,n}^2\underline{\eta}_i}
-\frac{\bar{\eta}_{i,d}}{2\underline{g}_{i,n}
\underline{\eta}_i^2}\right)$ and $\left| {{\varsigma_{i,q}}} \right| \le \bar{\varsigma}_{i,q}$ over $\Omega_v$ are exploited,  for $q=1,\cdots,n-1$.

By conducting a similar analysis to the proof of \emph{Theorem} 2, the results in \emph{Theorem} 1 can be easily demonstrated.
$\hfill{\blacksquare}$
\end{appendices}

\bibliographystyle{IEEEtran}
\bibliography{neuroadaptive}
\end{document}